\documentclass[twocolumn,showpacs,amsmath,nofootinbib,amssymb,epsfig]{revtex4}

\usepackage{graphicx}
\usepackage{dcolumn}
\usepackage{bm}

\newcommand{\Eqn}[1]{&\hspace{-0.2em}#1\hspace{-0.2em}&}

\begin{document}
\title{Energy conditions in $f(R,G)$ gravity}

\author{K. Atazadeh }
\email{atazadeh@azaruniv.ac.ir}\affiliation{Department of Physics, Azarbaijan Shahid Madani University , Tabriz, 53714-161 Iran\\Research Institute for Astronomy and Astrophysics of Maragha (RIAAM),
Maragha 55134-441, Iran}
\author{F. Darabi}
\email{f.darabi@azaruniv.ac.ir}\affiliation{Department of Physics, Azarbaijan Shahid Madani University , Tabriz, 53714-161 Iran\\Research Institute for Astronomy and Astrophysics of Maragha (RIAAM),
Maragha 55134-441, Iran}

\date{\today}

\begin{abstract}

Modified gravity is one of the most favorable candidates for explaining the current accelerating expansion of the Universe. In this regard, we study the viability of an alternative gravitational theory, namely $f(R,G)$, by imposing energy conditions. We consider two forms of $f(R,G)$, commonly discussed
in the literature, which account for the stability of cosmological solutions. We construct the inequalities obtained by energy conditions and specifically apply the weak energy condition using the recent estimated values of the Hubble, deceleration, jerk and snap parameters to probe the viability of the above-mentioned forms of $f(R, G)$.
\end{abstract}


\maketitle

\section{Introduction}

As is well known in general relativity, the energy conditions are often required in the proofs of various important theorems about black holes, such as no hair theorem or laws of black hole thermodynamics \cite{hawkingellis}.
The energy conditions are systematically obtained when one refers to the Raychaudhuri equation, in which the interesting property  of gravity is presented by the positivity condition $R_{\mu\nu}k^\mu k^\nu \geq 0$, while $R_{\mu\nu}$ and $k^\mu$ are Ricci tensor and any null vector, respectively. In Einstein's theory of general relativity (GR) this geometrical condition is equal to a matter condition, called null energy condition $T_{\mu\nu}k^\mu k^\nu \geq 0$, where $T_{\mu\nu}$ is the stress energy tensor.
Especially, under the weak energy condition (WEC), the local energy density is positive and also implies that $T_{\mu\nu}U^\mu U^\nu \geq 0$, where $U^\mu$ is a general timelike vector. For instance, in a perfect fluid case it is necessary to have
\begin{equation}\label{pf}
T_{\mu\nu}=(\rho+p)U^\mu U^\nu+pg_{\mu\nu},
\end{equation}
we have $\rho>0$ and $\rho+p\geq0$. As a result of continuity, the WEC in fact shows the null energy condition (NEC), $T_{\mu\nu}k^\mu k^\nu \geq 0$~\cite{hawkingellis}.\\
The energy conditions have been widely studied in in the context of modified gravity, such as $f(R)$, $f(G)$, $f(R,T)$ and $f(T)$ gravity \cite{energyconditions,Santos:2007bs}. The motivation for the study of modified theories of gravity is clear and
strong. As we know, Einstein's theory of General Relativity can not explain the late-time accelerated expansion of the Universe, unless an unknown dark energy element is introduced in the field equations \cite{expansion}. In the context of GR, the attempts to explain the observed speed-up of the Universe, have introduced some modifications of Einstein-Hilbert action by adopting a general function of the scalar curvature in the gravitational Lagrangian density as $f(R)$ \cite{fRgravity}. The motivation for this procedure consists of the analysis of strong gravitational fields near the curvature singularities and considering the consistent candidates of a fundamental theory of quantum gravity. According to string/M-theory predictions,  scalar field couplings with the Gauss-Bonnet invariant $G$ play an important role in the appearance of non-singular early time cosmologies. These impetuses have been studied to explain the late-time acceleration of the Universe \cite{Nojiri:2005vv,modGB1,modGB2}.

Recently, a new generalized modified Gauss-Bonnet gravity, whose action
contains a general function of $R$ and $G$ as $f(R,G)$, has attracted considerable attention \cite{F(GR)-gravity}. Besides its stability, this attention is due to its ability to describe the present acceleration of the universe as well as the phantom divide line crossing and transition from acceleration to deceleration phases.
Two specific models of $f(R,G)$ gravity were constructed to account for the late-time cosmic acceleration \cite{Nojiri:2007bt}, and the respective constraints of the parameters of the models were also analyzed in \cite{Nojiri:2007bt}. So, in order to proceed along with the interests
of these models, in the present work we shall consider two forms of $f(R,G)$
introduced in \cite{Nojiri:2007bt}.
In Ref.\cite{Nojiri:2007bt} the authors have studied the stability of de Sitter and power-law solutions in $f(R,G)$ gravity and have shown that gravitational action plays a very important role in the stability of the solutions both
of which depending on the form of the $f(R,G)$ theory and the parameters of the model. In this context, we further consider the constraints imposed by the energy conditions and verify whether the parameter range of the proposed models considered in \cite{Nojiri:2007bt} are consistent with the energy conditions. More specifically, we define generalized energy conditions for $f(R,G)$ modified theories of gravity, and consider their realization for flat Friedmann cosmological models. In particular, we analyze whether the weak energy condition is satisfied by particular choices of $f(R,G)$ which were advocated in Refs. \cite{Nojiri:2007bt}.\\

In Section \ref{ref:II}, we introduce the gravitational field equations for the $f(R,G)$ gravity. In Section \ref{ref:III}, we obtain the inequalities corresponding to the energy conditions. In Section \ref{ref:IV}, we consider two specific forms of $f(R,G)$, and analyze the constraints resulting from the energy conditions. The paper ends with a brief conclusions in Section \ref{ref:conclusion}. We use the units $c= G_{N}=1$.

\section{Field equations of $f(R,G)$ modified gravity}\label{ref:II}

Let us start by writing the most general action for modified Gauss-Bonnet gravity
\begin{equation}
S=\frac{1}{2\kappa}\int d^4x \sqrt{-g}f(R,G)+S_M(g^{\mu\nu},\psi)\,,
   \label{mdGBaction}
\end{equation}
where $S_M(g^{\mu\nu},\psi)$ is the matter action, and $f(R,G)$ is a function of the Ricci scalar and Gauss-Bonnet invariant defined by
\begin{equation}
G\equiv
R^2-4R_{\alpha\beta}R^{\alpha\beta}+R_{\alpha\beta\rho\sigma}
R^{\alpha\beta\rho\sigma}\,.
   \label{GBinvariant}
\end{equation}
Variation of the action (\ref{mdGBaction}) with respect to the metric provides the following gravitational field equation
\begin{equation}\label{eom}
R_{\mu \nu}-\tfrac{1}{2} g_{\mu \nu}R=\kappa\,T^{(\mathrm{mat})}_{\mu \nu}+\Sigma_{\mu \nu},
\end{equation}
where $\Sigma_{\mu \nu}$ is defined by
\begin{widetext}
\begin{eqnarray}
\Sigma_{\mu \nu}&=\nabla_\mu \nabla_\nu f_{R}-g_{\mu \nu} \Box f_{R}+2R \nabla_\mu \nabla_\nu f_{G}-2g_{\mu \nu} R \Box f_{G}-4R_\mu^{~\lambda} \nabla_\lambda \nabla_\nu f_{G}-4R_\nu^{~\lambda} \nabla_\lambda \nabla_\mu f_{G}
+4R_{\mu \nu} \Box f_{G} \nonumber \\
&\qquad+4 g_{\mu \nu} R^{\alpha \beta} \nabla_\alpha \nabla_\beta f_{G}+4R_{\mu \alpha \beta \nu} \nabla^\alpha \nabla^\beta f_{G}-\tfrac{1}{2}\,g_{\mu \nu} V +(1-f_{R})\,\bigl(R_{\mu \nu}-\tfrac{1}{2} g_{\mu \nu}R\bigr). \label{effective-energy-momentum}
\end{eqnarray}
\end{widetext}
Note that
\begin{equation}
  \label{eq:def1}
  f_{R}\equiv\frac{\partial f(R,G)}{\partial _{R}}\,,\qquad f_{G}\equiv \frac{\partial f(R,G)}{\partial_{G}},
\end{equation}
$V\equiv f_{R}R+f_{G} G-f(R,G)$, and $T^{(\mathrm{mat})}_{\mu \nu}$ is the stress energy tensor describing the ordinary matter.

Now, we consider the flat FRW metric
\begin{equation}
ds^{2}=-dt^{2}+a^{2}(t)(d{x}^{2}+d{y}^{2}+d{z}^{2})\,,
\label{metric}
\end{equation}
where $a(t)$ is the scale factor. In the FRW background with a perfect fluid equation of state for ordinary matter, the field equations for $f(R,G)$ gravity are given by
\begin{eqnarray}\label{6}
f_R\dot{H}&=&-\frac{\kappa}{2}(p^{(m)}+\rho^{(m)})+{1\over
2}(H\dot{f_R}-\ddot{f_R}+4H^3\dot{f_G}-8H\dot{H}\dot{f_G}\nonumber \\
&&-4H^2\ddot{f_G}),\\
f_R H^2&=&{\kappa\over 3}\rho^{(m)}+{1\over
6}(f_RR-f-6H\dot{f_R}+Gf_G-24H^3\dot{f_G})\nonumber.
\end{eqnarray}
where $\rho$ and $p$ are the energy density and pressure of ordinary matter, respectively, and the overdot denotes a derivative with respect to the time coordinate, $t$.
In addition , we have
\begin{eqnarray}
R = 6 \left(2H^{2}+\dot H \right)\,,
\label{eq:R} \\
G = 24H^{2} \left( H^{2}+\dot H \right)\,, \label{eq:G}
\end{eqnarray}
and the gravitational field equations may be rewritten in the following form
\begin{equation}
\rho_{\mathrm{eff}}=\frac{3}{\kappa}H^{2}\,,
\quad
p_{\mathrm{eff}}=-\frac{1}{\kappa} \left( 2\dot H+3H^{2} \right)\,,
\label{GutenTag}
\end{equation}
where $\rho_{\mathrm{eff}}$ and $p_{\mathrm{eff}}$ are
the effective energy density and pressure, respectively, defined by
\begin{widetext}
\begin{eqnarray}
\rho_{\mathrm{eff}} \Eqn{=} {1\over f_R}\left[\rho^{(m)}+{1\over
2\kappa}(f_RR-f-6H\dot{f_R}+Gf_G-24H^3\dot{f_G})\right]\,, \label{eq:rho-eff-1}
\\
p_{\mathrm{eff}} \Eqn{=} {1\over f_R}\left[p^{(m)}+{1\over
\kappa}\left(2H\dot{f_R}+\ddot{f}_{R}+8H^3\dot{f_G}+8H\dot{H}\dot{f_G}+4H^{2}\ddot{f}_{G}-\frac{1}{2}(Rf_R+Gf_G-f)\right)\right]
\,.
\label{eq:p-eff-1}
\end{eqnarray}
Combining the above equations, we obtain the following useful relationship
\begin{eqnarray}\label{eq:NEC}
\nonumber
\rho_{\mathrm{eff}}+p_{\mathrm{eff}} &=& {1\over f_R}\left[\rho^{(m)}+p^{(m)} + \frac{1}{\kappa}\left(-H\dot{f}_R-4H^{3}\dot{f}_G+\ddot{f}_{R}+8H\dot{H}\dot{f}_G+4H^{2}\ddot{f}_G\right)\right],
\end{eqnarray}
which will be used throughout the paper.
\end{widetext}

\section{Energy Conditions}\label{ref:III}

In general, the energy conditions emanate when one studies the Raychaudhuri equation given by
\begin{equation}
\label{Raych}
\frac{d\theta}{d\tau}= - \frac{1}{2}\,\theta^2 -
\sigma_{\alpha\beta}\sigma^{\alpha\beta} + \omega_{\alpha\beta}\omega^{\alpha\beta}
- R_{\alpha\beta}k^{\alpha}k^{\beta} \;,
\end{equation}
where  $\theta\,$, $\sigma^{\alpha\beta}$ and $\omega_{\alpha\beta}$ are the expansion, shear and rotation, respectively, associated with the congruences defined by the null vector field $k^{\alpha}$.

From Raychaudhury equation it is seen that for spatial shear tensor with $\sigma^2 \equiv \sigma_{\alpha\beta}\sigma^{\alpha\beta}\geq 0$, and for any hypersurface orthogonal congruences, which imposes $\omega_{\alpha\beta} \equiv 0$, the condition for attractive gravity, namely ${d\theta}/{d\tau}<0$ reduces to $R_{\mu\nu}k^{\mu}k^{\nu}\geq 0$. In general relativity, using the Einstein field equations one can rewrite the above condition in terms of the stress-energy tensor given by $T_{\mu\nu}k^\mu k^\nu \ge 0$. However, in any other theory of gravity such as $f(R,G)$, one should know how to replace $R_{\mu\nu}$ in terms of $T_{\mu\nu}$, using the corresponding field equations.

Equation (\ref{eom}) may be written in the following effective form
\begin{equation}\label{field:eq2}
G_{\mu\nu}\equiv R_{\mu\nu}-\frac{1}{2}R\,g_{\mu\nu}= T^{{\rm
eff}}_{\mu\nu} \,,
\end{equation}
where the effective energy momentum tensor is given by

\begin{eqnarray}
T^{{\rm eff}}_{\mu\nu}
&=& \kappa T^{(\mathrm{mat})}_{\mu \nu}
+\Sigma_{\mu \nu}.
\end{eqnarray}

The positivity condition $R_{\mu\nu}k^\mu k^\nu \geq 0$, through the modified gravitational field equation (\ref{field:eq2}), supplies the following form for the null energy condition
\begin{equation}
T^{{\rm eff}}_{\mu\nu} k^\mu k^\nu\geq 0.
\end{equation}
Moreover, it is plausible to impose the condition $T^{({\rm mat})}_{\mu\nu} k^\mu k^\nu\ge 0$ for ordinary matter, because it implies that the energy density of ordinary matter is positive in all local frames of references.

To deduce the energy conditions in the context of $f(R,G)$ modified gravity, we use the modified (effective) gravitational field equations and obtain
the energy conditions as follows
\begin{equation}\label{NEC}
{\rm NEC}\Leftrightarrow \rho_{\mathrm{eff}} +p_{\mathrm{eff}} \geq 0 \,,
\end{equation}
\begin{equation}\label{WEC}
{\rm WEC} \Leftrightarrow \rho_{\mathrm{eff}} \geq 0 \; {\rm and} \; \rho_{\mathrm{eff}} +p_{\mathrm{eff}} \geq 0 \,,
\end{equation}
\begin{equation}\label{SEC}
{\rm SEC} \Leftrightarrow \rho_{\mathrm{eff}} +3p_{\mathrm{eff}} \geq 0 \; {\rm and} \; \rho_{\mathrm{eff}} +p_{\mathrm{eff}} \geq 0 \,,
\end{equation}
\begin{equation}\label{DEC}
{\rm DEC} \Leftrightarrow \rho_{\mathrm{eff}} \geq 0 \; {\rm and} \; \rho_{\mathrm{eff}} \pm p_{\mathrm{eff}} \geq 0 \,,
\end{equation}
here we have advocated the symbols NEC, WEC, SEC and DEC for the null, weak, strong and dominant energy conditions, respectively.

To continue, in analogy with the standard mechanics we introduce velocity, acceleration, jerk and snap in the cosmological context. Thus, in addition to the Hubble parameter $H=\dot{a}/a$, the deceleration, jerk, and snap parameters are defined by
\begin{equation}
q=-\frac{1}{H^2}\frac{\ddot{a}}{a}\;,~~~~~ j=\frac{1}{H^3}\frac{\dddot{a}}{a}\;,
~~~~ {\rm{and}} ~~~~ s=\frac{1}{H^4}\frac{\ddddot{a}}{a}\;,
\end{equation}
respectively; in terms of which, we may consider the following definitions
\begin{eqnarray}\label{HH}
&\dot{H}=-H^2(1+q)\;, \\
&\ddot{H}=H^3(j+3q+2)\;, \\
&\dddot{H}=H^4(s-2j-5q-3)\;,\label{HHH}
\end{eqnarray}
respectively.
By using (\ref{HH})-(\ref{HHH}) one may rewrite equations (\ref{eq:rho-eff-1}) and (\ref{eq:p-eff-1}) in the explicit forms, as follows
\begin{widetext}
\begin{eqnarray}\label{266}
\nonumber
\rho_{\mathrm{eff}}&=&\frac{1}{f_R}\left[\rho^{(m)}+\frac{1}{2
   \kappa }(-24 H^4 q f_G-576 H^8 f_{\text{GG}} \left(j+2 q^2+3 q\right)-1728 H^6 f_{\text{GR}} \left(j+q^2+q-1\right)-\right.\nonumber\\&&\left.36 H^4 f_{\text{RR}} (j-q-2)+6 H^2 (1-q) f_R-f)\right],
\end{eqnarray}
\begin{eqnarray}\label{267}
\nonumber
p_{\mathrm{eff}}&=&\frac{1}{f_R}\left[p^{(m)}+\frac{1}{\kappa }[576 \left(2 q^2+3 q+j\right)^2 \left(4 f_{\text{GGG}} H^2+f_{\text{GGR}}\right) H^{10}+24 (-2 j (3 q+2)-\right.\nonumber\\&&\left.q (2 q (q+6)+5)+s+3) \left(4 f_{\text{GG}}
   H^2+f_{\text{GR}}\right) H^6+144 (j-q-2) \left(2 q^2+3 q+j\right) \left(8 f_{\text{GGR}} H^2+2 f_{\text{RRG}}\right) H^6+\right.\nonumber\\&&\left.36 (j-q-2)^2 \left(4 f_{\text{RRG}}
   H^2+f_{\text{RRR}}\right) H^6+24 \left(2 q^2+3 q+j\right) \left(8 f_{\text{GG}} H^3-8 (q+1) f_{\text{GR}} H^3+2 f_{\text{GR}} H\right) H^5+\right.\nonumber\\&&\left.6 \left(4 q^2+15 q+2 j+s+9\right)
   \left(4 f_{\text{GG}} H^2+f_{\text{RR}}\right) H^4+6 (j-q-2) \left(-8 (q+1) f_{\text{GR}} H^3+8 f_{\text{GR}} H^3+2 f_{\text{RR}} H\right) H^3+\right.\nonumber\\&&\left.\frac{1}{2} \left(24 q f_G
   H^4-6 (1-q) f_R H^2+f\right)]\right],
\end{eqnarray}
where $\rho^{(m)}$ and $p^{(m)}$ are the matter energy density and pressure, respectively.
To derive the above equations we have replaced for $\dot{f}_{R}$, $\dot{f}_{G}$, $\ddot{f}_{G}$ and $\ddot{f}_{R}$ in equation (\ref{eq:rho-eff-1}) and (\ref{eq:p-eff-1}), for example as
 \begin{eqnarray}
\dot{f}_{R}(R,G)=f_{RR}\dot{R}+f_{RG}\dot{G}=6(4H\dot{H}+\ddot{H})f_{RR}+(96H^{3}\dot{H}+48H\dot{H}^{2}+24H^{2}\ddot{H})f_{RG}.
\end{eqnarray}

Using these definitions, the energy conditions (\ref{NEC})-(\ref{DEC}) take on the following respective forms
%\vspace{1cm}
\begin{eqnarray}\label{26}
\nonumber
{\rm NEC}:\\  \nonumber\qquad \rho_{\textrm{eff}}+p_{\textrm{eff}}&=&
p^{(m)}+\rho^{(m)}+\frac{1}{\kappa}\left[24 H^6 (-2 j (3 q+2)-q (2 q (q+6)+5)+s+3) \left(4 H^2 f_{\text{GG}}+f_{\text{GR}}\right)+\right.\nonumber\\&&\left.24 H^5 (j+q (2 q+3)) \left(8 H^3 f_{\text{GG}}-2 H f_{\text{GR}} \left(4 H^2
   (q+1)-1\right)\right)-288 H^8 f_{\text{GG}} (j+q (2 q+3))+\right.\nonumber\\&&\left.6 H^4 \left(2 j+4 q^2+15 q+s+9\right) \left(4 H^2 f_{\text{GG}}+f_{\text{RR}}\right)+576 H^{10} (j+q (2
   q+3))^2 \left(4 H^2 f_{\text{GGG}}+f_{\text{GGR}}\right)+\right.\nonumber\\&&\left.288 H^6 (j-q-2) (j+q (2 q+3)) \left(4 H^2 f_{\text{GGR}}+f_{\text{RRG}}\right)-864 H^6 f_{\text{GR}}
   \left(j+q^2+q-1\right)-\right.\nonumber\\&&\left.12 H^3 (j-q-2) \left(4 H^3 q f_{\text{GR}}-H f_{\text{RR}}\right)-18 H^4 f_{\text{RR}} (j-q-2)+\right.\nonumber\\&&\left.36 H^6 (-j+q+2)^2 \left(4 H^2
   f_{\text{RRG}}+f_{\text{RRR}}\right)\right]\geq 0\,,
\end{eqnarray}
\begin{eqnarray}\label{27}
{\rm WEC}:\\  \nonumber\qquad
\rho_{\textrm{eff}}&=&\rho^{(m)}+\frac{1}{2\kappa}\left[-f(R,G)-24 H^4 q f_G-576 H^8 f_{\text{GG}} \left(j+2 q^2+3 q\right)-24^{2} H^6 f_{\text{GR}} \left(j+q^2+q-1\right)\right.\\\nonumber&&\left.-36 H^4 f_{\text{RR}} (j-q-2)+6 H^2 (1-q) f_R\right]\geq 0, ~~ \qquad\rho_{\textrm{eff}}+p_{\textrm{eff}}\geq 0\,,
\end{eqnarray}
\begin{eqnarray}\label{28}\nonumber
{\rm SEC}: \qquad \rho_{\textrm{eff}}+3p_{\textrm{eff}}&=&\rho^{(m)}+3p^{(m)}+\frac{1}{\kappa}[3(0.5(24 H^4 q f_G+6 H^2 (q-1) f_R+f)+24 H^6 (-2 j (3 q+2)-\nonumber\\&& q (2 q (q+6)+5)+s+3) (4 H^2 f_{\text{GG}}+f_{\text{GR}})+24 H^5 (j+q
   (2 q+3)) (8 H^3 f_{\text{GG}}-\nonumber\\&&2 H f_{\text{GR}} (4 H^2 (q+1)-1))+6 H^4 (2 j+q (4 q+15)+s+9) (4 H^2
   f_{\text{GG}}+f_{\text{RR}})+\nonumber\\&&576 H^{10} (j+q (2 q+3))^2(4 H^2 f_{\text{GGG}}+f_{\text{GGR}})+288 H^6 (j-q-2) (j+\nonumber\\&&q (2 q+3)) (4 H^2
   f_{\text{GGR}}+f_{\text{RRG}})-12 H^4 (j-q-2) (4 H^2 q f_{\text{GR}}-\nonumber\\&&f_{\text{RR}})+36 H^6 (-j+q+2)^2 (4 H^2
   f_{\text{RRG}}+f_{\text{RRR}}))+\frac{1}{2} (6 H^2 (-4 H^2 q f_G-\nonumber\\&&96 H^6 f_{\text{GG}} (j+q (2 q+3))-288 H^4 f_{\text{GR}}
   (j+q^2+q-1)+\nonumber\\&&6 H^2 f_{\text{RR}} (-j+q+2)-(q-1) f_R)-f)]\geq 0  ~,\qquad\rho_{\textrm{eff}}+p_{\textrm{eff}}\geq 0 \,,
\end{eqnarray}
\begin{eqnarray}\label{29}
{\rm DEC}:\\ \nonumber \qquad\rho_{\textrm{eff}}-p_{\textrm{eff}}&=&\rho^{(m)}-p^{(m)}-\frac{1}{\kappa}\left[24 H^4 q f_G-24 H^6 (-2 j (3 q+2)-q (2 q (q+6)+5)+s+3) \left(4 H^2 f_{\text{GG}}+f_{\text{GR}}\right)\right.\\\nonumber&&\left.-24 H^5 (j+q (2 q+3)) \left(8 H^3 f_{\text{GG}}-2 H
   f_{\text{GR}} \left(4 H^2 (q+1)-1\right)\right)-\right.\\\nonumber&&\left.288 H^8 f_{\text{GG}} (j+q (2 q+3))-6 H^4 (2 j+q (4 q+15)+s+9) \left(4 H^2 f_{\text{GG}}+f_{\text{RR}}\right)-\right.\\\nonumber&&\left.576
   H^{10} (j+q (2 q+3))^2 \left(4 H^2 f_{\text{GGG}}+f_{\text{GGR}}\right)-\right.\\\nonumber&&\left.288 H^6 (j-q-2) (j+q (2 q+3)) \left(4 H^2 f_{\text{GGR}}+f_{\text{RRG}}\right)-864 H^6
   f_{\text{GR}} \left(j+q^2+q-1\right)+\right.\\\nonumber&&\left.12 H^4 (j-q-2) \left(4 H^2 q f_{\text{GR}}-f_{\text{RR}}\right)+18 H^4 f_{\text{RR}} (-j+q+2)-6 H^2 (q-1) f_R-\right.\\\nonumber&&\left.36 H^6
   (-j+q+2)^2 \left(4 H^2 f_{\text{RRG}}+f_{\text{RRR}}\right)-f\right]\geq 0\,,
~~~~ \rho_{\textrm{eff}}+p_{\textrm{eff}}\geq 0,~~~~ \rho_{\textrm{eff}}\geq 0 \,.
\end{eqnarray}
\end{widetext}

%%%%%%%%%%%%%%%%%%%%%%%%%%%%%%%%%%%%%%%%
\section{CONSTRAINTS on $f(R,G)$ GRAVITY}\label{ref:IV}
%%%%%%%%%%%%%%%%%%%%%%%%%%%%%%%%%%%%%%%%

In this section, we consider the viable $f(R,G)$ modified theories of gravity which were used in \cite{Nojiri:2007bt} to study the stability of cosmological solutions. Stability of power-law solutions for $f(R,G)$ models were given by expressions (\ref{uno}) and (\ref{terzo}), bellow. In the cases where no fluids are considered (vacuum) it is seen that stability of the cosmological solutions can be achieved by appropriate choices of the parameters space
\cite{Nojiri:2007bt}. However, in this section we consider the
constraints imposed by the energy conditions and verify the consistency
between the parameters ranges of the models considered in \cite{Nojiri:2007bt} and the energy conditions in $f(R,G)$ flat Friedman cosmological models
obtained here as (\ref{26})-(\ref{29}).

In this regard, we consider two classes of viable $f(R,G)$ given by \cite{Nojiri:2007bt}
\begin{eqnarray}
f_{1}(R,G) \Eqn{=} \mu R^{\beta} G^{\gamma}\,,
\label{uno} \\
f_{2}(R,G) \Eqn{=} k_1R+k_2R^{n}G^{m}\,,\label{terzo}
\end{eqnarray}
where $k_{1}$, $k_{2}$, $\mu$, $n$, $m$, $\beta$, and $\gamma$
are constants and $n, \beta$ are assumed to be positive. The Ricci scalar and Gauss-Bonnet invariant, defined in equations (\ref{eq:R}) and (\ref{eq:G}) in terms of the Hubble and deceleration parameters, can also be expressed as
\begin{equation}
R=6H^{2}(1-q)\,,
\end{equation}
\begin{equation}
G=-24H^{4}q \,,
\end{equation}
respectively.

Since the inequalities (\ref{26})-(\ref{29}) imposed by the energy conditions in $f(R,G)$ gravity are so lengthly, for simplicity we only consider the WEC in the following analysis. Moreover, we take the following observed values for the deceleration, jerk and snap parameters \cite{Rap,Poplawski:2006ew}:
$s_0=-0.22^{+0.21}_{-0.19}$,  $j_0=2.16^{+0.81}_{-0.75}$, and $q_0=-0.81\pm 0.14$.

%%%%%%%%%%%%%%%%%%%%%%%%%%%%%%%%%%%%%%%%
\subsection{$ f_{1}(R,G) =  \mu R^{\beta} G^{\gamma}$}
%%%%%%%%%%%%%%%%%%%%%%%%%%%%%%%%%%%%%%%%

To impose energy conditions on $f_1(R,G)$ gravity, for simplicity we consider the vacuum, i.e., $\rho^{(m)} =p^{(m)}=0$. Hence, the WEC constraints for the vacuum, $\rho_{\textrm{eff}}^{(v)}\geq 0$ and $\rho_{\textrm{eff}}^{(v)}+p_{\textrm{eff}}^{(v)}\geq 0$, are given respectively
by
\begin{widetext}
\begin{eqnarray}\label{NEC1a}
&&2^{\beta +3 \gamma -1} 3^{\beta +\gamma } \left(-H^2 (q-1)\right)^{\beta } \left(-H^4 q\right)^{\gamma } \left[\frac{(-j+q+2) (\beta -1) \beta }{(q-1)^2}-\frac{12
   \left(q^2+q+j-1\right) \gamma  \beta }{(q-1) q}\right.\\\nonumber&&\left.+\beta -\frac{(j+q (2 q+3)) (\gamma -1) \gamma }{q^2}+\gamma -1\right] \mu
\geq 0 \,,
\end{eqnarray}
\begin{eqnarray}\label{NEC1b}
&&2^{\beta +3 \gamma -3} 3^{\beta +\gamma -1} \left(-H^2 (q-1)\right)^{\beta } \left(-H^4 q\right)^{\gamma } \left[-\frac{4 (\beta -1) \beta  (q (\beta +\gamma -2)-\gamma )
   (-j+q+2)^2}{(q-1)^3 q}+\right.\\\nonumber&&\left.\frac{12 (\beta -1) \beta  (-j+q+2)}{(q-1)^2}-\frac{8 \beta  (\beta -q \gamma +\gamma -1) (-j+q+2)}{(q-1)^2}+\right.\\\nonumber&&\left.\frac{8 (j+q (2 q+3)) \beta  \gamma
   (-\gamma +q (\beta +\gamma -2)+1) (-j+q+2)}{H^2 (q-1)^2 q^2}-\frac{144 \left(q^2+q+j-1\right) \beta  \gamma }{(q-1) q}-\right.\\\nonumber&&\left.\frac{12 (j+q (2 q+3)) (\gamma -1) \gamma
   }{q^2}-\frac{8 (j+q (2 q+3)) \gamma  \left(q \left(\left(4 H^2 (q+1)-1\right) \beta -\gamma +1\right)+\gamma -1\right)}{(q-1) q^2}+\right.\\\nonumber&&\left.\frac{(2 j+q (4 q+15)+s+9) \left((\gamma
   -1) \gamma  (q-1)^2+4 H^2 q^2 (\beta -1) \beta \right)}{H^2 (q-1)^2 q^2}-\right.\\\nonumber&&\left.\frac{4 (j+q (2 q+3))^2 (\gamma -1) \gamma  (-\gamma +q (\beta +\gamma -2)+2)}{(q-1) q^3}-\right.\\\nonumber&&\left.\frac{4 (j
   (6 q+4)+q (2 q (q+6)+5)-s-3) \gamma  (-\gamma +q (\beta +\gamma -1)+1)}{(q-1) q^2}\right] \mu \geq 0\,.
\end{eqnarray}
\begin{figure}[ht]
  \centering
  \includegraphics[width=3in]{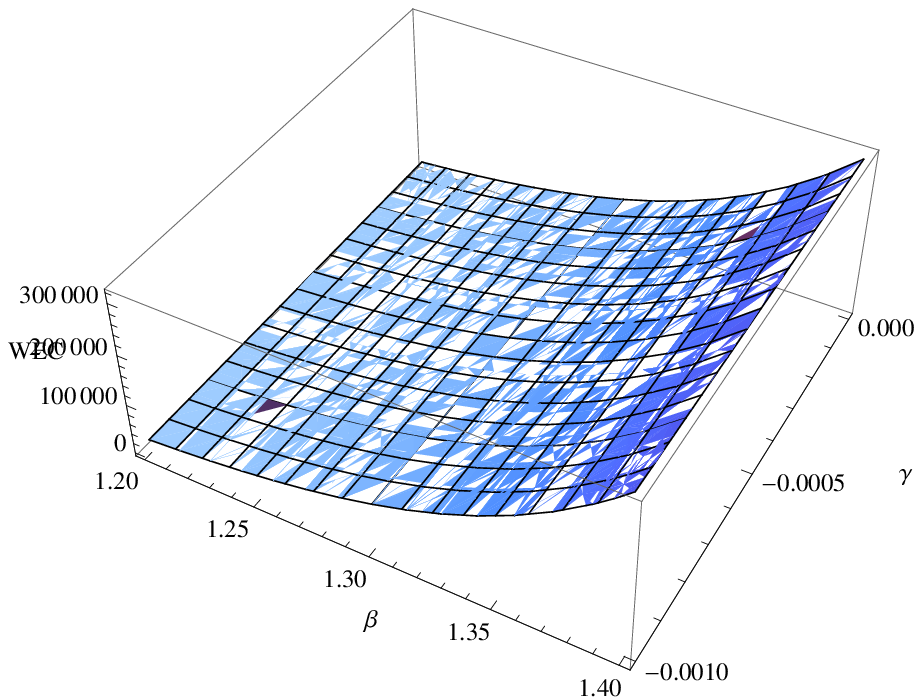}~
   \includegraphics[width=3in]{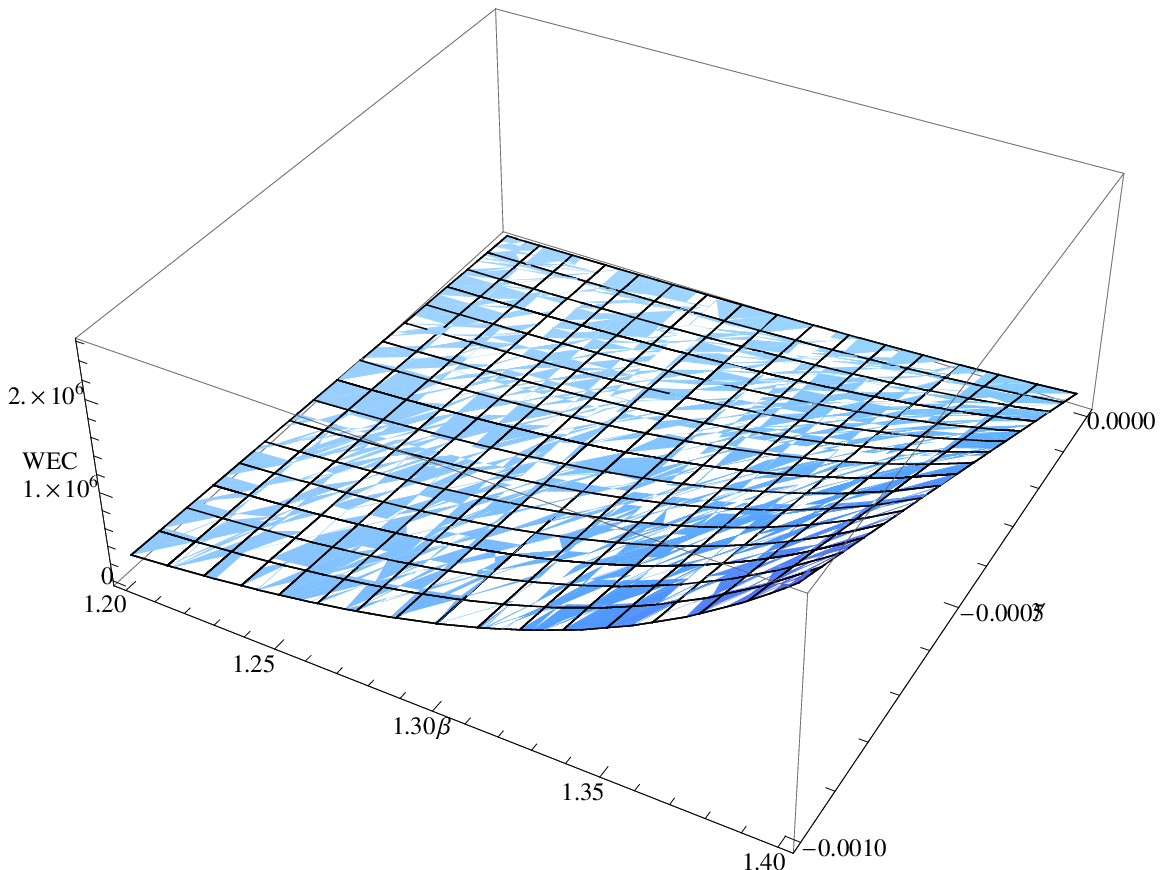}
  \caption{Plots of the weak energy condition for the specific form of $ f_{1}(R,G) = \mu R^{\beta} G^{\gamma}$. The left and right plots correspond respectively to $\rho_{\textrm{eff}}^{(v)}\geq 0$ and $\rho_{\textrm{eff}}^{(v)}+p_{\textrm{eff}}^{(v)}\geq 0$. The positivity requirement of the weak energy condition is satisfied in the plots for the parameter range considered.}
  \label{fig:WEC1a}
\end{figure}
\end{widetext}

\begin{figure*}[ht]
  \centering
  \includegraphics[width=2.6in]{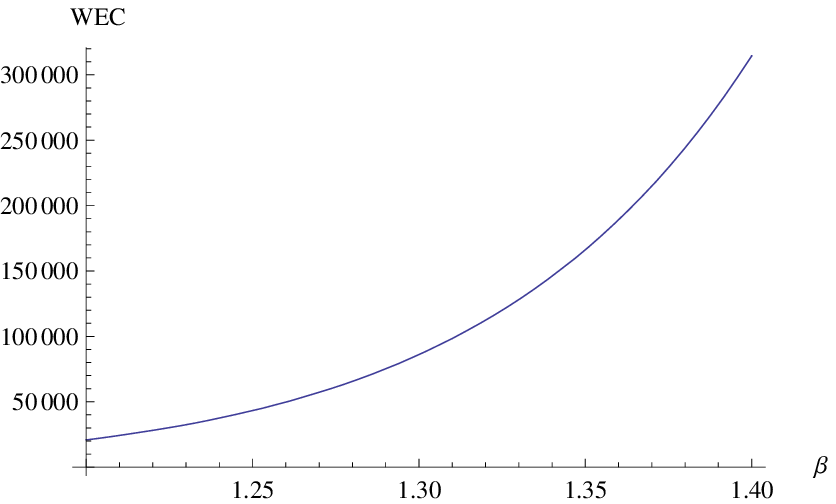}
   \includegraphics[width=2.6in]{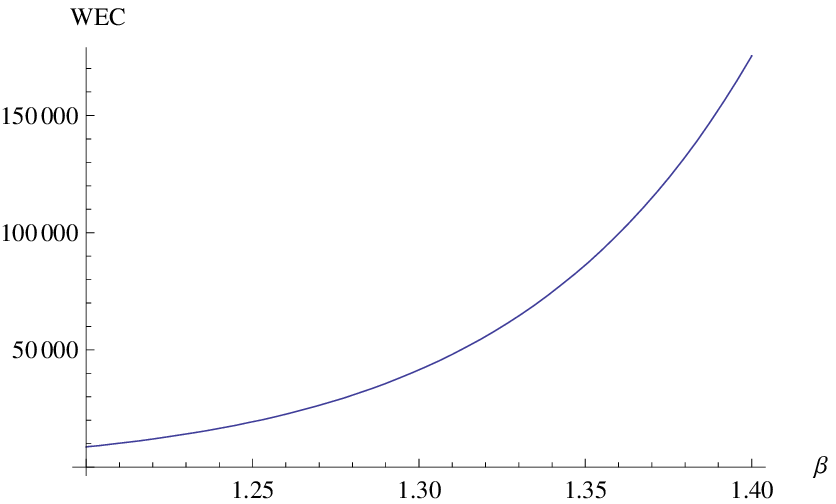}
  \caption{The plots show the weak energy condition for $f_{1}(R) = \mu R^{\beta}$. The left plot corresponds to $\rho_{\textrm{eff}}^{(v)}\geq 0$; the right plot corresponds to $\rho_{\textrm{eff}}^{(v)}+p_{\textrm{eff}}^{(v)}\geq 0$. }
  \label{fig:WECB2}
\end{figure*}

The exact analytical expressions for the parameter ranges of the constants $\beta$ and $\gamma$ can not be explicitly obtained because the constraints provided by the inequalities (\ref{NEC1a})-(\ref{NEC1b}) are so complex, so we consider specific values for some of the parameters. In \cite{Nojiri:2007bt} the authors have shown that the vacuum solutions are stable for $\beta>1$ and $\gamma<\frac{1}{8}(5-4\beta)$.
To verify this result from the energy condition's point of view, we plot the WEC as a function of $\beta$ and $\gamma$, as is depicted in Fig. \ref{fig:WEC1a}.
It is seen that in the left (right) figure, the weak energy condition $\rho_{\textrm{eff}}^{(v)}\geq 0$  ($\rho_{\textrm{eff}}^{(v)}+p_{\textrm{eff}}^{(v)}\geq 0$) is satisfied for the parameter ranges $\beta\geq1$ and $\gamma<0$, in the specific form of  $f_{1}(R,G)$ gravity by equation (\ref{uno}).

In order to compare with the particular case $f(R)$, we put $\gamma=0$ in equation  (\ref{uno}). We find that equation (\ref{uno}) can be reduced to $f_1(R)=\mu R^{\beta}$, which is a class of $f(R)$-gravity.
The energy conditions of such theories have been studied in \cite{Santos:2007bs}. Therefore, considering $\gamma=0$ in the current model  the WEC constraints, i.e., $\rho_{\textrm{eff}}^{(v)}\geq 0$ and $\rho_{\textrm{eff}}^{(v)}+p_{\textrm{eff}}^{(v)}\geq 0$, are given respectively
by
\begin{eqnarray}\label{273}
 &&\frac{1}{(q-1)^2}[2^{\beta -1} 3^{\beta } \left(-H^2 (q-1)\right)^{\beta } (\beta -1) \left((q-1)^2\right.\\\nonumber&&\left.+(-j+q+2) \beta \right)]\geq 0\,,  \end{eqnarray}
 \begin{eqnarray}\label{2733}
&&-\frac{1}{(q-1)^3}[6^{\beta -1} \left(-H^2 (q-1)\right)^{\beta } (\beta -1) \beta  \left((\beta -2) j^2\right.\\\nonumber&&\left.+3 (q+3) j-2 (q+2) \beta  j+s-q (2 q (2 q+7)+s+3)+\right.\\\nonumber&&\left.q (q+4) \beta +4 \beta
+3\right)]\geq 0\,.
\end{eqnarray}
For more understanding the above relations, we have plotted them in terms of $\beta$. Obviously  Fig. 2 shows  that the WEC is valuable for $\beta\geq1$, and this is in agreement with result obtained in \cite{Santos:2007bs}

%%%%%%%%%%%%%%%%%%%%%%%%%%%%%%%%%%%%%%%%
\subsection{
$f_{2}(R,G) = k_1R+k_2R^{n}G^{m}$}
%%%%%%%%%%%%%%%%%%%%%%%%%%%%%%%%%%%%%%%%

\begin{figure*}[ht]
  \centering
  \includegraphics[width=3in]{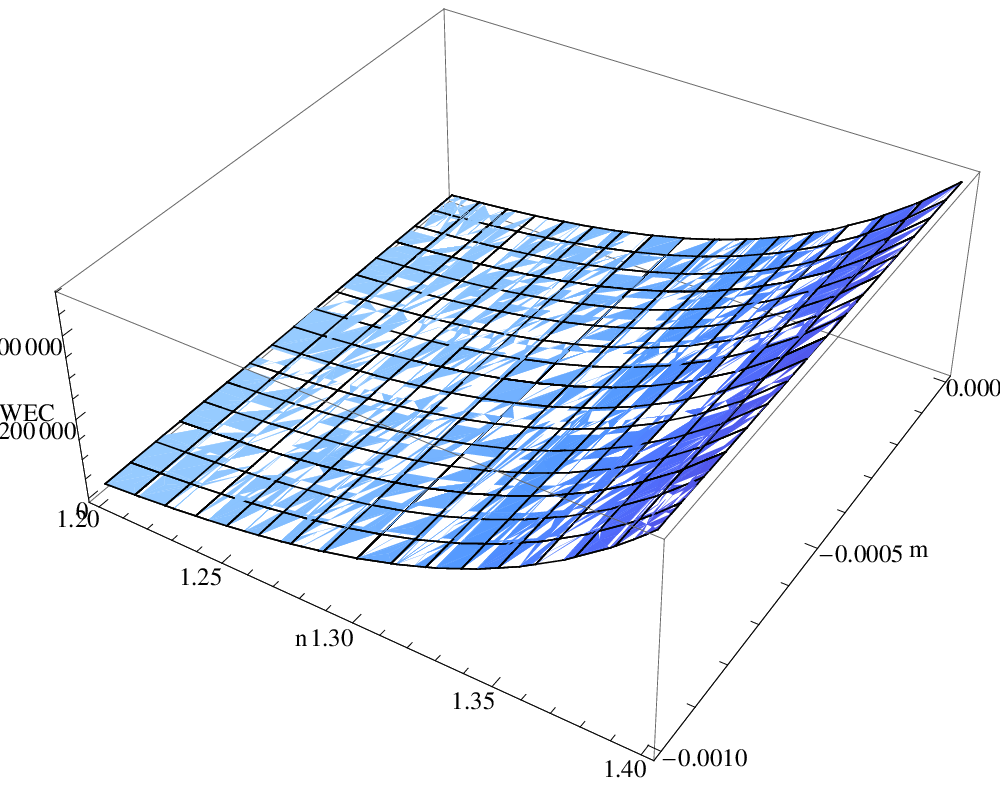}~~~
   \includegraphics[width=3in]{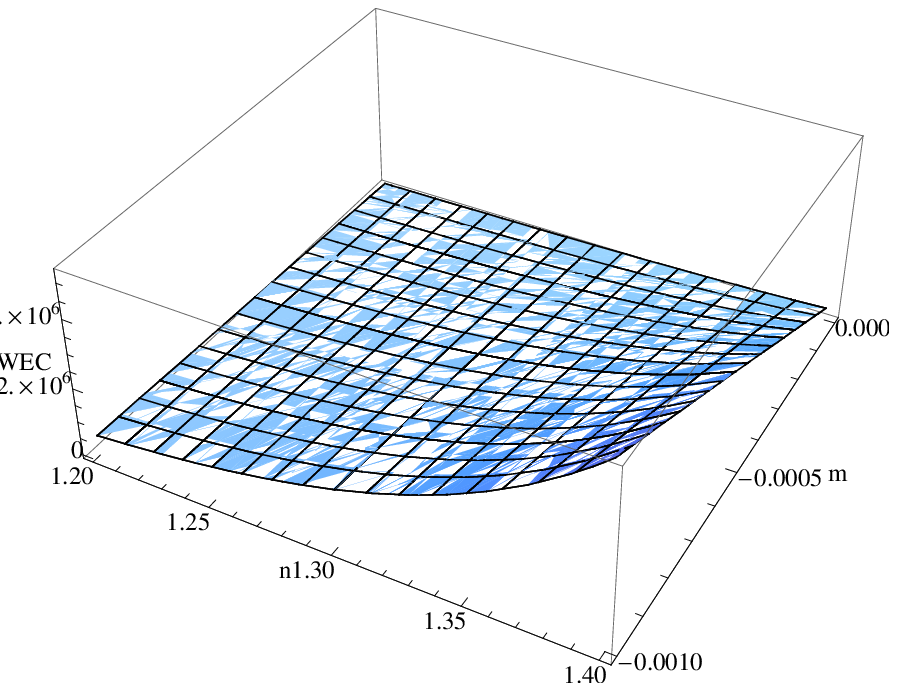}
  \caption{The plots clarify the weak energy condition for $f_{2}(R,G) = k_1R+k_2R^{n}G^{m}$. The left plot corresponds to $\rho_{\textrm{eff}}^{(v)}\geq 0$; the right plot corresponds to $\rho_{\textrm{eff}}^{(v)}+p_{\textrm{eff}}^{(v)}\geq 0$. }
  \label{fig:WECB2}
\end{figure*}

In this case, as in the previous one, we consider the vacuum, i.e., $\rho^{(m)} =p^{(m)}=0$. This modified gravity model accounts for the stability of the cosmological solutions for late-time cosmic acceleration, and the stability of solutions
depends on the values of the coupling constants $k_1$ and $k_2$. In fact, for $k_1 >2(\frac{1152}{25})^{3/2} k_2$ the perturbations grow exponentially, and the de Sitter solution becomes unstable. For other values of $k_1$ and $k_2$ than mentioned above, the perturbations behave as damped oscillations decaying to zero, hence the solution becomes stable. To verify whether the parameter range of the stability of solutions are consistent with the energy conditions, we again consider the weak energy condition.
\\
For the form of $f_2(R,G)$ considered by equation (\ref{terzo}), the WEC constraints, i.e., $\rho_{\textrm{eff}}^{(v)}\geq 0$ and $\rho_{\textrm{eff}}^{(v)}+p_{\textrm{eff}}^{(v)}\geq 0$, are given by
\begin{widetext}
\begin{eqnarray}\label{NEC2a}
&& -\frac{2^{3 m+n-1} 3^{m+n}k_2}{{(q-1)^2 q^2}}[ \left(-H^2 (q-1)\right)^n \left(-H^4 q\right)^m \left(j \left(m^2 (q-1)^2+m ((12 n-1) q+1) (q-1)+(n-1) n q^2\right)\right.\\\nonumber&&\left.+q \left(m^2 (2 q+3) (q-1)^2+3 m
   \left(-q^2+4 n \left(q^2+q-1\right)+1\right) (q-1)-(n-1) q \left((q-1)^2+n (q+2)\right)\right)\right) ]\geq 0 \,,
\end{eqnarray}
\begin{eqnarray}\label{NEC2b}
&&2^{3 m+n-3} 3^{m+n-1}(-H^2 (q-1))^n (-H^4 q)^m  k_2\left[-\frac{4 (n-1) n ((m+n-2) q-m) (-j+q+2)^2}{(q-1)^3 q}+\right.\\\nonumber&&\left.\frac{8 m n (-m+(m+n-2) q+1) (j+q (2 q+3))
   (-j+q+2)}{H^2 (q-1)^2 q^2}+\frac{12 (n-1) n (-j+q+2)}{(q-1)^2}-\right.\\\nonumber&&\left.\frac{4 (m-1) m (-m+(m+n-2) q+2) (j+q (2 q+3))^2}{(q-1) q^3}+\frac{8 n (j-q-2) (-q
   m+m+n-1)}{(q-1)^2}-\right.\\\nonumber&&\left.\frac{144 m n (q^2+q+j-1)}{(q-1) q}-\frac{12 (m-1) m (j+q (2 q+3))}{q^2}-\frac{8 m (j+q (2 q+3)) (-q m+m+q+n q (4 H^2
   (q+1)-1)-1)}{(q-1) q^2}-\right.\\\nonumber&&\left.\frac{4 m (-m+(m+n-1) q+1) (j (6 q+4)+q (2 q (q+6)+5)-s-3)}{(q-1) q^2}+\right.\\\nonumber&&\left.\frac{(m^2 (q-1)^2-m (q-1)^2+4 H^2 (n-1) n q^2) (2 j+q
   (4 q+15)+s+9)}{H^2 (q-1)^2 q^2}\right]\geq 0\,,
\end{eqnarray}
\end{widetext}
respectively.

It is seen that the weak energy condition dose not depend on $k_1$. Therefore, it turns out that WEC do not interfere with the stability conditions. As in the previous example, considering the complex constraints provided by the inequalities (\ref{NEC2a})-(\ref{NEC2b}), finding exact analytical expressions for the parameter ranges of the constants $m$ and $n$ is not
an easy task. Hence, we consider specific values for the parameters to find WEC condition as a function of the parameters $m$ and $n$.
It is seen in the left (right) figure.3 that the weak energy condition $\rho_{\textrm{eff}}^{(v)}\geq 0$ ($\rho_{\textrm{eff}}^{(v)}+p_{\textrm{eff}}^{(v)}\geq 0$) is satisfied for the parameter ranges $n\geq1$ and $m <0$, in the specific form of  $f_{2}(R,G)$ gravity by equation (\ref{terzo}).

Again, in order to compare with the particular case $f(R)$ we put $m=0$ in equation (\ref{terzo}) to obtain  $f_2(R)=k_{1}R+k_{2}R^{n}$.
The energy conditions of such theories have been studied in \cite{Santos:2007bs}. Thus, considering $m=0$ in our model, the WEC constraints, i.e., $\rho_{\textrm{eff}}^{(v)}\geq 0$ and $\rho_{\textrm{eff}}^{(v)}+p_{\textrm{eff}}^{(v)}\geq 0$, are given respectively
by
\begin{eqnarray}\label{274}
&&\frac{1}{(q-1)^2}[2^{n-1} 3^n (n-1) \left(-H^2 (q-1)\right)^n \left((q-1)^2\right.\\\nonumber&&\left.+n (-j+q+2)\right)]\geq 0\,,
\end{eqnarray}
\begin{eqnarray}\label{2744}
&& -\frac{1}{(q-1)^3}[6^{n-1} (n^{2}-n)  \left(-H^2 (q-1)\right)^n \left((n-2) \right.\\\nonumber&&\left.j^2+(3 (q+3)-2 n (q+2)) j+n (q+2)^2+\right.\\\nonumber&&\left.s-q (2 q (2 q+7)+s+3)+3\right)]\geq 0\,.
\end{eqnarray}
We have plotted equations (\ref{274}) and (\ref{2744})  in terms of $\beta$. Fig. 4 shows that the WEC is valuable for $n\geq1$, and this is in agreement with the results obtained in \cite{Santos:2007bs}

\begin{figure*}[ht]
  \centering
  \includegraphics[width=2.6in]{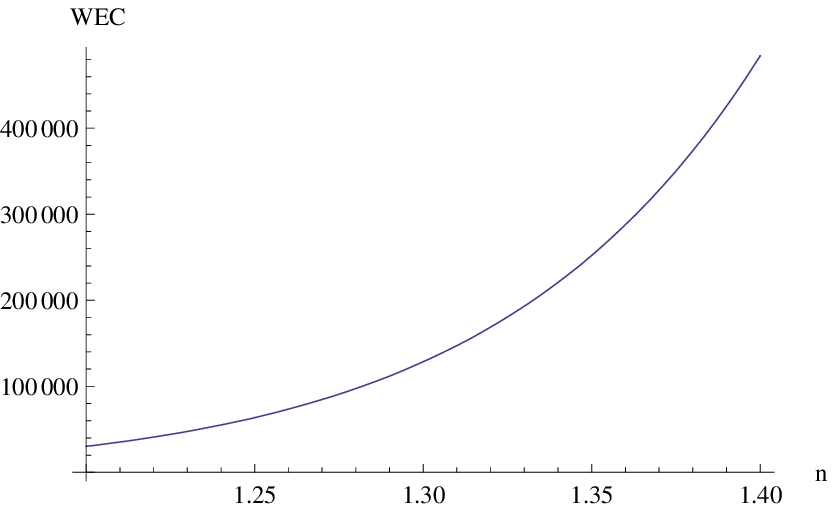}
   \includegraphics[width=2.6in]{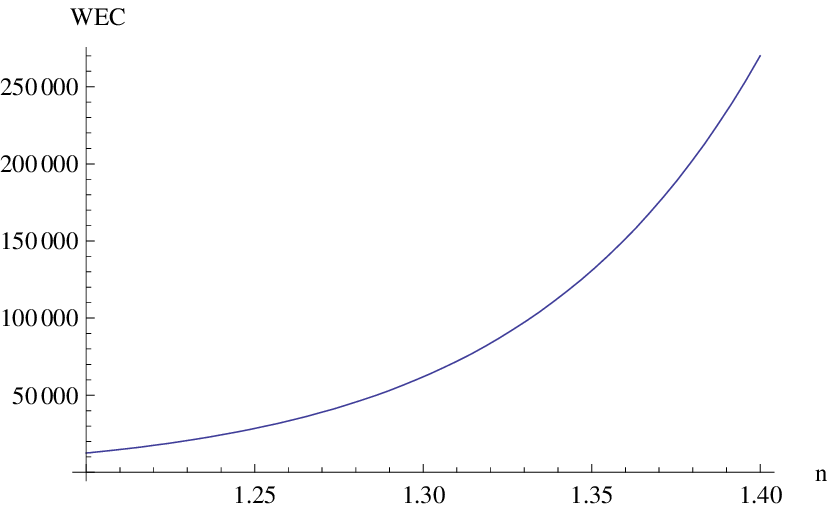}
  \caption{The plots show the weak energy condition for $f_{2}(R) = k_1R+k_2R^{n}$. The left plot corresponds to $\rho_{\textrm{eff}}^{(v)}\geq 0$; the right plot corresponds to $\rho_{\textrm{eff}}^{(v)}+p_{\textrm{eff}}^{(v)}\geq 0$. }
  \label{fig:WECB2}
\end{figure*}

To complete our discussions, we have plotted SEC and DEC in the Figs. 5 and 6 for two suggested $f(R,G)$ models (\ref{uno}) and (\ref{terzo}). From these figurers it can be seen that in both models SEC holds but DEC does not. This is in agreement with the results obtained previously in the study of energy conditions in $f(R)$ gravity \cite{Santos:2007bs}.

Note that for simplicity, we have examined the vacuum case $p^{(m)} = \rho^{(m)} = 0$. If we add regular matter to our models, the general results of the paper will not change, because with no loss of generality we may always add a positive energy density or pressure of matter satisfying the WEC to the vacuum case such that $\rho_{\textrm{eff}}\geq 0$, $\rho_{\textrm{eff}}+p_{\textrm{eff}}\geq 0$.

%%%%%%%%%%%%%%%%%%%%%%
\section{Conclusions}\label{ref:conclusion}
%%%%%%%%%%%%%%%%%%%%%%

In this paper, we have studied the viability of another alternative gravitational theory, namely, $f(R,G)$ gravity. We have considered two realistic models of $f(R,G)$, analyzed in the literature, accounting for the late-time cosmic acceleration and the stability of the cosmological solutions \cite{Nojiri:2007bt}. We have obtained the general inequalities by imposing the energy conditions. To
be specific and for simplicity, we have focused on the weak energy condition and used the recent observational data of the Hubble, deceleration, jerk and snap parameters. We have shown the consistency of the above-mentioned forms of $f(R,G)$ with the weak energy condition.

 We have just examined the vacuum case for which $p^{(m)} = \rho^{(m)} = 0$. Actually  this is not a physically relevant case bearing in mind that the universe contains matter. This simplification, however,
does not change the general results of the paper, if we add matter to our models. This is  because we can add a positive energy density or pressure of regular matter satisfying the WEC ($\rho^{(m)}\geq0$ and $\rho^{(m)}+p^{(m)}\geq0$) to the vacuum case ($\rho^{(v)}\geq0$ and $\rho^{(v)}+p^{(v)}\geq0$) such that $\rho_{\textrm{eff}}\geq 0$, $\rho_{\textrm{eff}}+p_{\textrm{eff}}\geq 0$, where $\rho_{\textrm{eff}}=\rho^{(m)}+\rho^{(v)}$ and $p_{\textrm{eff}}=p^{(m)}+p^{(v)}$.
\begin{figure*}[ht]
  \centering
  \includegraphics[width=3in]{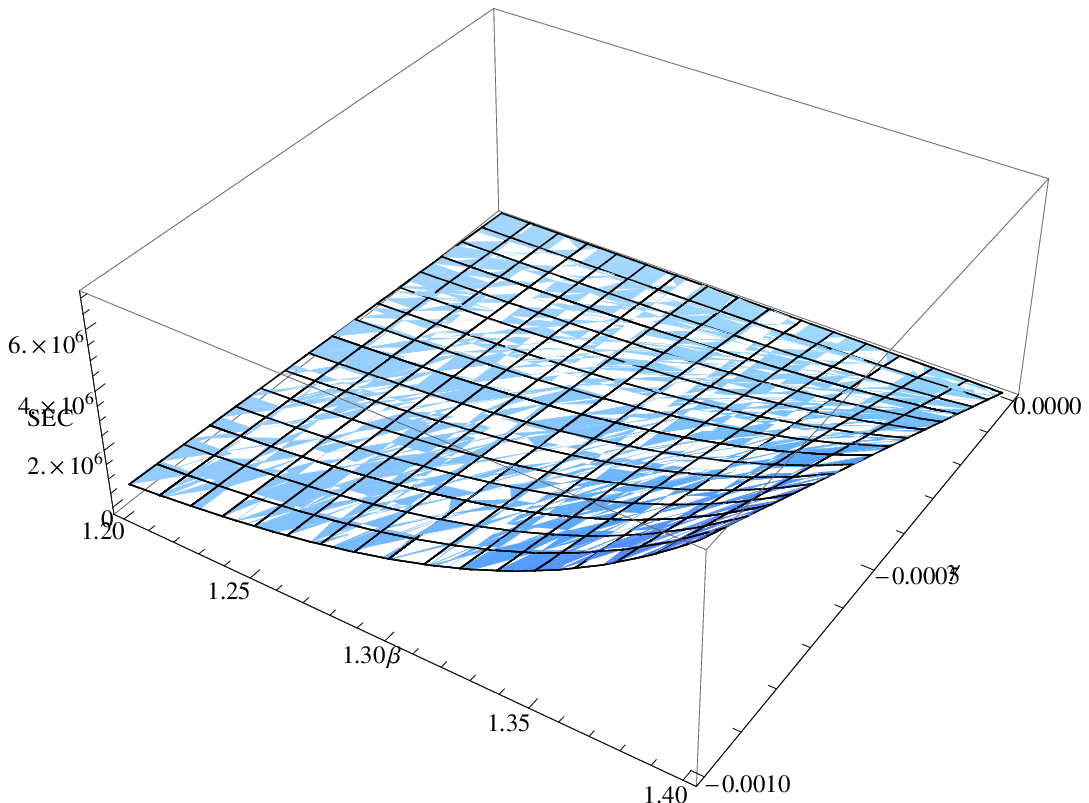}
   \includegraphics[width=3in]{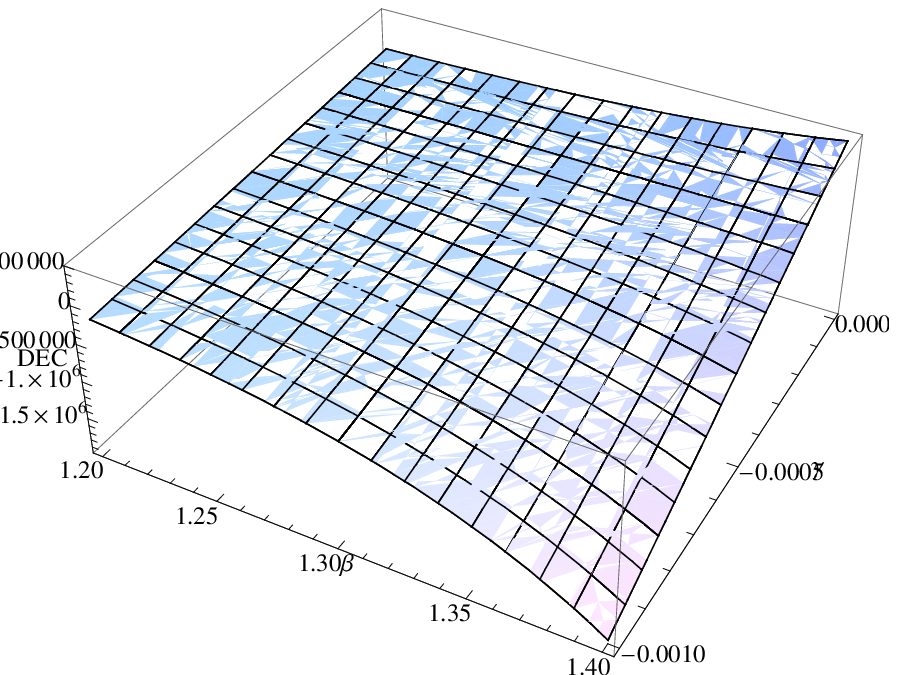}
  \caption{The plots present the strong and dominant energy conditions for  $f_{1}(R,G) = \mu R^{\beta}G^{\gamma}$. The left plot corresponds to SEC; the right plot corresponds to DEC. }
  \label{fig:WECB2}
\end{figure*}
\begin{figure*}[ht]
  \centering
  \includegraphics[width=3in]{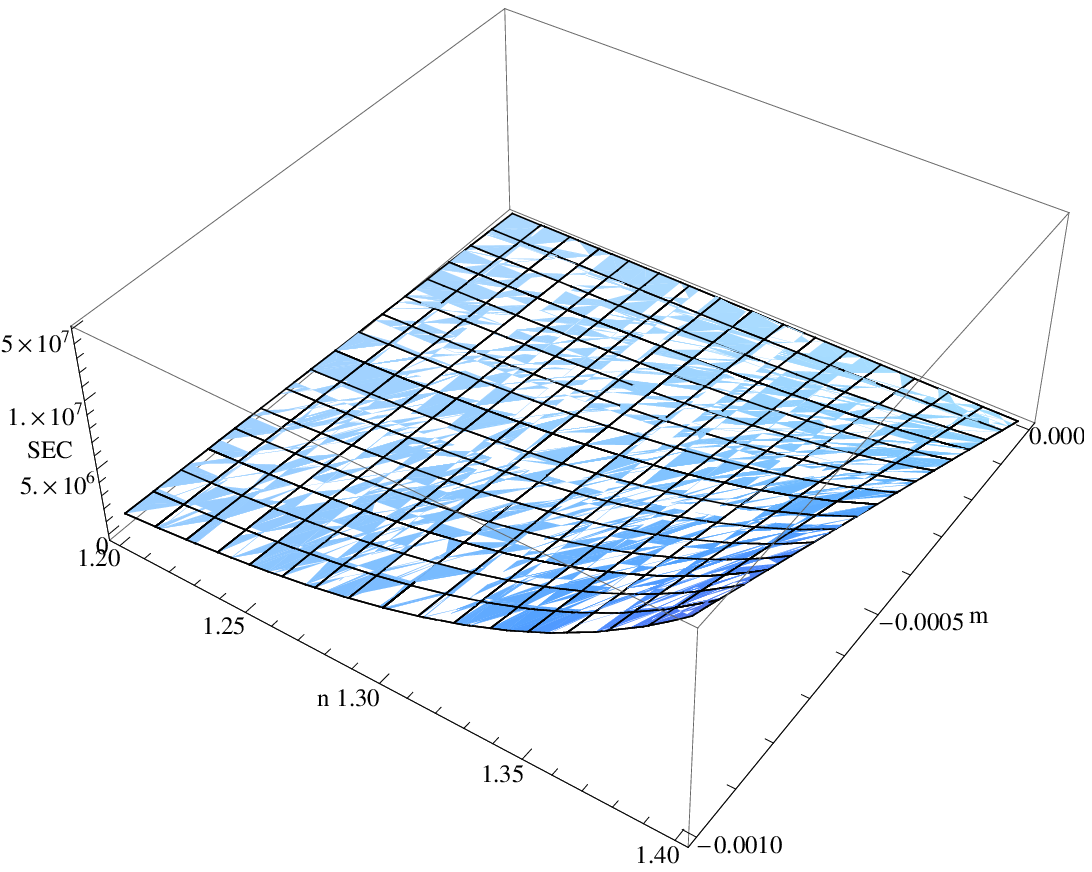}
   \includegraphics[width=3in]{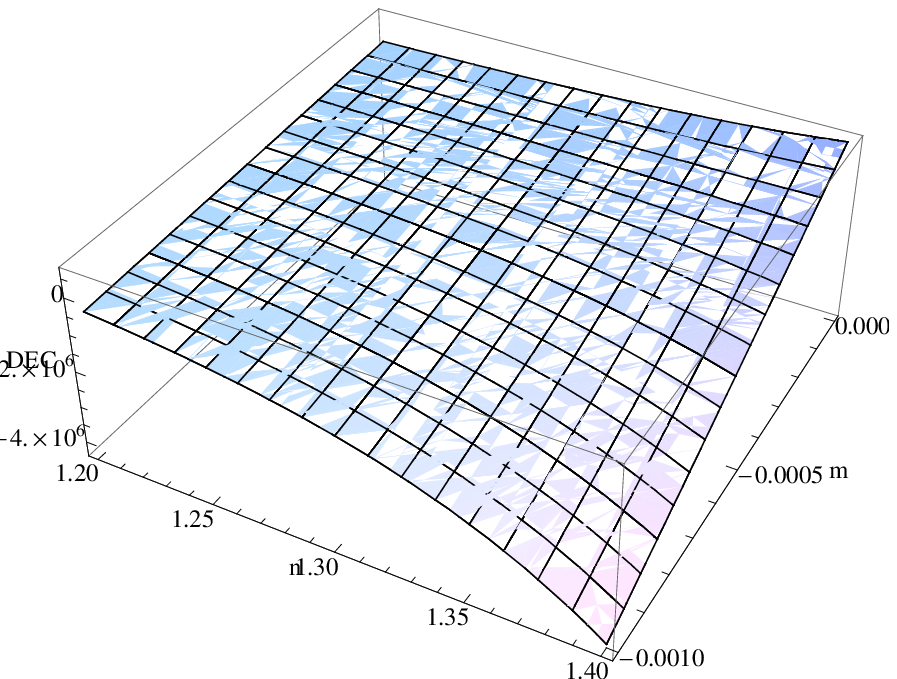}
  \caption{The plots present the strong and dominant  energy conditions for $f_{2}(R,G) = k_1R+k_2R^{n}G^{m}$. The left plot corresponds to SEC; the right plot corresponds to DEC. }
  \label{fig:WECB2}
\end{figure*}

\section*{Acknowledgments}
This work has been supported financially by Research Institute for Astronomy and Astrophysics of
Maragha (RIAAM) under research project No.1/2782-58.


\begin{thebibliography}{99}
\bibitem{hawkingellis}
S. W. Hawking and G.F.R. Ellis, {\em The Large Scale Structure of Spacetime},(Cambridge University Press, England, (1973).
\bibitem{energyconditions}
J.H. Kung, Phys. Rev. D {\bf 52} (1995) 6922; Phys. Rev. D {\bf 53} (1996) 3017;
%
M. Visser, Science {\bf 276} (1997) 88; Phys. Rev. D {\bf 56} (1997) 7578;
J. Santos and J.S. Alcaniz, Phys. Lett. B {\bf 619} (2005) 11;
J. Santos, J. S. Alcaniz and M. J. Rebou\c{c}as, Phys. Rev. D {\bf 74} (2006) 067301;
S.E. Perez Bergliaffa, Phys. Lett. B {\bf 642} (2006) 311;
J. Santos, J. S. Alcaniz, N. Pires and M. J. Rebou\c{c}as, Phys. Rev. D {\bf 75} (2007) 083523;
%
J. Santos, J. S. Alcaniz, M.~J.~Reboucas and N.~Pires,
    Phys.\ Rev.\  D {\bf 76} (2007) 043519;
    \\
 K. Atazadeh, A. Khaleghi, H. R. Sepangi and Y. Tavakoli, Int. J. Mod. Phys. D {\bf18} (2009) 1101;
\\
M. Sharif and M. Zubair, J. Phys. Soc. Jpn. {\bf82} (2013) 014002.

\bibitem{Santos:2007bs}
  J. Santos, J.~S. Alcaniz, M. J.~Reboucas and F. C.~Carvalho,
   Phys.\ Rev.\  D {\bf 76} (2007) 083513;
    \\N. M. Garc\'{\i}a, T. Harko, F. S. N. Lobo and J. P. Mimoso, Phys. Rev. D {\bf 83} (2011) 104032;\\
N. M. Garc\'{\i}a, T. Harko, F. S. N. Lobo, J. P. Mimoso, J. Phys. Conf. Ser. 314 (2011) 012060;
\\
Di Liu and M. J. Reboucas, Phys.Rev. D {\bf86} (2012) 083515; \\
F. G. Alvarenga, M. J. S. Houndjo, A. V. Monwanou and Jean B. Chabi Orou, J. of Mod. Phys. {\bf4} (2013) 130.

\bibitem{expansion}
 A.~G. Riess {\it et al.}, Astron.\ J.\  {\bf 116} (1998) 1009;
S.~Perlmutter {\it et al.}, Astrophys.\ J.\  {\bf 517} (1999) 565;
S. Perlmutter, M. S. Turner and M. White, Phys. Rev. Lett.
{\bf 83} (1999) 670;
A. Grant {\it et al}, Astrophys. J. {\bf 560} (2001) 49;
C. L. Bennett {\it et al}, Astrophys. J.
Suppl. {\bf 148} (2003) 1;
A.~G. Riess {\it et al.}, Astrophys.\ J.\  {\bf 607} (2004) 665;
E.~J. Copeland, M.~Sami and S. Tsujikawa,
   Int.\ J.\ Mod.\ Phys.\  D {\bf 15} (2006) 1753.
\bibitem{fRgravity}
H. A. Buchdahl, Mon. Not. Roy. Astron. Soc. {\bf 150} (1970) 1;
A.~A. Starobinsky, Phys.\ Lett.\  B {\bf 91} (1980) 99;
%
R. Kerner, Gen. Rel. Grav. {\bf 14} (1982) 453;
J. P. Duruisseau, R. Kerner and P. Eysseric, Gen.
Rel. Grav. {\bf 15} (1983) 797;
J. D. Barrow and A. C. Ottewill, J. Phys. A:
Math. Gen. {\bf 16} (1983) 2757;
%
L.~M.~Sokolowski,
  Class.\ Quant.\ Grav.\  {\bf 24} (2007) 3391;
  %
  %
C.~G. B\"ohmer, L.~Hollenstein and F. S.~N.~Lobo, Phys.\
Rev.\  D {\bf 76} (2007) 084005;
  G.~J.~Olmo, Phys.\ Rev.\ \textbf{D75} (2007) 023511;
%
S. Capozziello, V.~F. Cardone and A. Troisi,
JCAP {\bf0608}, 001 (2006); S. Capozziello, V. F. Cardone and
A. Troisi, Mon. Not. R. Astron. Soc. \textbf{375} (2007) 1423;
%
A. Borowiec, W. Godlowski and M. Szydlowski,
Int. J. Geom. Meth. Mod. Phys. \textbf{4} (2007) 183;
%
C.~F. Martins and P.~Salucci,
  Mon.\ Not.\ Roy.\ Astron.\ Soc.\  {\bf 381} (2007) 1103;
%
S. Carloni, P. K.~S.~Dunsby and A.~Troisi,
  Phys.\ Rev.\  D {\bf 77} (2008) 024024;
  %
K.~N. Ananda, S. Carloni and P.~K.~S.~Dunsby,
    Phys.\ Rev.\  D {\bf 77} (2008) 024033;
G.~Cognola, M.~Gastaldi and S.~Zerbini,
    Int.\ J.\ Theor.\ Phys.\  {\bf 47} (2008) 898;
A.~De Felice and S. Tsujikawa, Living Rev.\ Rel.\  {\bf 13} (2010) 3;
\bibitem{Nojiri:2005vv}
 S. Nojiri, S. D. Odintsov and M. Sasaki, Phys.\ Rev.\  D { \bf71} (2005) 123509.

\bibitem{modGB1}
S. Nojiri and S. D. Odintsov,  Int.\ J.\ Geom.\ Meth.\ Mod.\ Phys.,  {\bf4} (2007) 115;

\bibitem{modGB2}
G. Cognola, E. Elizalde, S. Nojiri, S. D. Odintsov and
S. Zerbini,
  Phys.\ Rev.\  D, { \bf73} (2006)  084007;
 %
S. Nojiri, S. D. Odintsov O. G. and Gorbunova, J.\ Phys.\
A: Math. Gen. { \bf39} (2006) 6627.

%%%%%
\bibitem{F(GR)-gravity}
 S. Nojiri, S. D. Odintsov and M. Sami,
  Phys.\ Rev.\  D {\bf 74} (2006) 046004 ;
 S. Nojiri and S.~D.~Odintsov,
  J.\ Phys.\ Conf.\ Ser.\  {\bf 66} (2007) 012005;
D. Bazeia, B. Carneiro da Cunha, R. Menezes and A.~Y.~Petrov,
    Phys.\ Lett.\  B {\bf 649} (2007) 445
%
G. Cognola, E. Elizalde, S. Nojiri, S. Odintsov and S. Zerbini,
 Phys.\ Rev.\  D {\bf 75} (2007) 086002;\
 B. Li, J.~D.~Barrow and D.~F.~Mota,
   Phys.\ Rev.\  D {\bf 76} (2007)044027 ;\
D. Bazeia, R.~Menezes and A.~Y.~Petrov,
    Eur.\ Phys.\ J.\  C {\bf 58} (2008) 171;\
   N. Goheer, R. Goswami, P.~K.~S.~Dunsby and K.~Ananda,
  Phys.\ Rev.\  D {\bf 79} (2009) 121301;\
 Phys.\ Lett.\  B {\bf 679} (2009) 302;\
 A. De Felice and S. Tsujikawa,
    Phys.\ Rev.\  D {\bf 80} (2009) 063516 ;
    M.~Mohseni,
    Phys.\ Lett.\  B {\bf 682} (2009) 89 ;
H. Mohseni Sadjadi, Europhys. Lett. {\bf92} (2010) 50014;
A. De Felice and S.~Tsujikawa,
  Phys.\ Lett.\  B {\bf 675} (2009) 1 ;\
 M.~Alimohammadi and A.~Ghalee,
 Phys.\ Rev.\  D {\bf 79} (2009) 063006 ;\
 C. G. Boehmer and F.~S.~N.~Lobo,
   Phys.\ Rev.\  D {\bf 79} (2009) 067504;\
 K.~Uddin, J.~E.~Lidsey and R.~Tavakol,
    Gen.\ Rel.\ Grav.\  {\bf 41} (2009) 2725;
   S.~Y.~Zhou, E.~J.~Copeland and P.~M.~Saffin,
  JCAP {\bf 0907} (2009) 009 ;\
 E. Elizalde, R. Myrzakulov, V. V. Obukhov and D. S\'{a}ez-G\'{o}mez, Class. Quant. Grav. {\bf27} (2010) 095007.

\bibitem{Nojiri:2007bt}
 \'{A}. de la Cruz-Dombriz and D. S\'{a}ez-G\'{o}mez, Class. Quant. Grav. {\bf29} (2012) 245014.

\bibitem{Rap}
D. Rapetti, S.W. Allen, M.A. Amin and R.D. Blandford,
Mont. Not. R. Soc. {\bf375} (2007) 1510.
\bibitem{Poplawski:2006ew}
  N. J. Poplawski,
    Class.\ Quant.\ Grav.\  {\bf 24} (2007) 3013.

\end{thebibliography}
\end{document}